\documentclass[aps,pre,twocolumn]{revtex4-1}
\usepackage[utf8]{inputenc}
\usepackage[T1]{fontenc}

\usepackage{amsmath}
\usepackage{amsfonts}
\usepackage{amssymb}
\usepackage{graphicx}
\usepackage{dsfont}

\usepackage[greek,english]{babel}
\newcommand\koppa{\text{\selectlanguage{greek}{\qoppa}}}

\usepackage{physics}

\usepackage{hyphenat}
\usepackage{hyperref}
\usepackage{xcolor}
\usepackage{xspace}

\newcommand\ie{\mbox{i.\,e.}\xspace}

\newcommand\Fig[1]{Fig.~\ref{#1}}
\renewcommand\Ref[1]{Ref.~\cite{#1}}
\newcommand\Refs[1]{Refs.~\cite{#1}}
\newcommand\Eq[1]{Eq.~\eqref{#1}}

\newcommand\duppercrit{\ensuremath{d_\mathrm{uc}}}
\newcommand\hc{\ensuremath{h_\mathrm{c}}}
\newcommand\Hamil{\mathcal H}
\newcommand\Observ{\mathcal O}
\newcommand{\eexp}{\mathrm{e}} 
\newcommand{\expect}[1]{\langle #1\rangle}
 
\newcommand\linchain{linear chain}
\newcommand\sqlattice{square lattice}

\renewcommand{\vec}[1]{\boldsymbol{\mathbf{#1}}}

\usepackage[normalem]{ulem}

\newcommand{\myZ}{Z}
\newcommand{\myTr}{\text{Tr}}
\newcommand{\mySn}{S_n}
\newcommand{\mySL}{S_\mathcal{L}}

\begin{document}

\title{Quantum-critical properties of the long-range transverse-field Ising model from quantum Monte Carlo simulations}
\author{Jan Alexander Koziol}
\affiliation{Lehrstuhl f\"ur Theoretische Physik I, Staudtstra{\ss}e 7, Universit\"at Erlangen-N\"urnberg, D-91058 Erlangen, Germany}
\author{Anja Langheld}
\affiliation{Lehrstuhl f\"ur Theoretische Physik I, Staudtstra{\ss}e 7, Universit\"at Erlangen-N\"urnberg, D-91058 Erlangen, Germany}
\author{Sebastian C. Kapfer}
\affiliation{Lehrstuhl f\"ur Theoretische Physik I, Staudtstra{\ss}e 7, Universit\"at Erlangen-N\"urnberg, D-91058 Erlangen, Germany}
\author{Kai Phillip Schmidt}
\affiliation{Lehrstuhl f\"ur Theoretische Physik I, Staudtstra{\ss}e 7, Universit\"at Erlangen-N\"urnberg, D-91058 Erlangen, Germany}

\begin{abstract}
    The quantum-critical properties of the transverse-field Ising model with algebraically decaying interactions 
    are investigated by means of stochastic series expansion quantum Monte Carlo,
    on both the one-dimensional linear chain and the two-dimensional square lattice.
	We extract the critical exponents $\nu$ and $\beta$ as a function of the decay exponent of the long-range interactions. For ferromagnetic Ising interactions, we resolve the limiting regimes known from field theory, ranging from the nearest-neighbor Ising to the long-range Gaussian universality classes, as well as the intermediate regime with continuously varying critical exponents. In the long-range Gaussian regime, we treat the effect of dangerous irrelevant variables on finite-size scaling forms. For antiferromagnetic and therefore competing Ising interactions, the stochastic series expansion algorithm displays growing auto-correlation times leading to a reduced performance. Nevertheless, our results are consistent with the nearest-neighbor Ising universality for all investigated interaction ranges both on the \linchain\ and the \sqlattice.
\end{abstract}

\maketitle


Currently quantum phase transitions in systems with long-range interactions receive a lot of interest, since they display remarkable quantum critical properties which become more and more accessible in experimental set-ups. In particular, Ising-type interactions with algebraically decaying coupling constants are well realized in quantum optical platforms like trapped Rydberg atoms \cite{Schauss2012} or cold ions \cite{Friedenauer2008,Kim2009,Kim2010,Edwards2010,Lanyon2011,Islam2011,Schneider2012,Britton2012,Knap2013,Islam2013,Jurcevic2014,Richerme2014,Bohnet2016,Yang2019}. Furthermore, long-range interactions can also play an essential role in understanding exotic collective phenomena of material properties in condensed matter physics, e.g., materials with dipolar interactions like the so-called spin-ice materials \cite{Castelnovo2008,Bramwell2001} or the dipolar ferromagnet LiHoF$_4$ \cite{Chakraborty2004,Bitko1996}. Long-range transverse-field Ising models are therefore one of the most paradigmatic systems to investigate quantum phenomena which originate from the long-range nature of the interaction.

The microscopic treatment of the long-range transverse-field Ising model (LRTFIM) is challenging and a variety of numerical methods were developed to tackle such systems. This includes high-order series expansions \cite{Coester2015,Fey2019,Adelhardt2020}, tensor-network approaches \cite{Vodola2015,Sun2017,Saadatmand2018,Vanderstraeten2018}, quantum Monte Carlo simulations \cite{Sandvik2003,Humeniuk2016,Humeniuk2018,Humeniuk2020}, and the functional renormalization group \cite{Defenu2017}.
The ferromagnetic LRTFIM is of particular interest, both in one and two dimensions.
From field theory, we expect the universality class of the quantum phase transition between the high-field polarized and the low-field symmetry broken phase to change as a function of the range of interactions,
from short-range to being described by a long-range Gaussian theory \cite{Fisher1972,Sak1973,Sak1977,Dutta2001,Defenu2017}.
A regime with continuously varying critical exponents is expected to connect those well-known limits.
Precise values for the critical exponents in this intermediate interaction regime are subject of current research. In particular, the transition to the nearest-neighbour criticality is only poorly understood. 
Further, long-range interactions can provide a path to explore the physics of phase transitions above the upper critical dimension in low-dimensional systems \cite{Luijten1997,Flores2015}.

The quantum critical properties of the antiferromagnetic LRTFIM are known to be distinct from its ferromagnetic counterpart. Only in the nearest-neigbor limit on bipartite lattices, there is an exact duality between the ferromagnetic and antiferromagnetic case and the quantum critical properties coincide. Otherwise, one has to distinguish on the one hand antiferromagnetic LRTFIMs on lattices with a strong degree of geometrical frustration in the nearest-neighbor limit like the triangular, kagome, or pyrochlore lattice and, on the other hand, antiferromagnetic long-range Ising interactions on bipartite lattices so that the nearest-neighbor limit is unfrustrated. In the first case the interplay of frustration and long-range interactions represents a formidable challenge. Rich phase diagrams are known to occur \cite{Humeniuk2016,Saadatmand2018, Fey2019, Koziol2019}, but a full understanding is still in its infancy. In the second case the antiferromagnetic long-range interactions induce competing interactions so that several studies indicate that the universality of the quantum phase transitions remains in the nearest-neighbor universality class in a large range of interactions \cite{Fey2016, Sun2017, Fey2019}. Deviations might only occur in the regime of ultra long-range interactions \cite{Koffel2012,Vodola2015}.

In this article, we focus on the zero-temperature limit of the LRTFIM.
We use the stochastic series expansion (SSE) quantum Monte Carlo (QMC) algorithm \cite{Sandvik2003} to simulate on the order of 1000 spins with either ferromagnetic or antiferromagnetic couplings, both on the \linchain\ and the \sqlattice. 
We demonstrate that critical properties such as the critical field values $\hc$, and the critical exponents $\beta$ and $\nu$ of the quantum phase transition can be extracted from our unbiased simulations by means of finite-size scaling.
For the ferromagnetic case, we extract the scaling power $\tilde{y}_r$ of the tuning parameter $r=|h/h_c-1|$ at the quantum phase transition which is normally associated with the critical exponent $\nu$. In the long-range Gaussian regime, this correspondence is modified by dangerous irrelevant variables. 
Inspired by the recently proposed Q-finite-size scaling method \cite{Berche2012,Kenna2014a,Kenna2014b,Flores2015} we provide a consistent way to calculate $\nu$ from $\tilde{y}_r$ for the LRTFIM on the \linchain\ and on the \sqlattice\ in the long-range Gaussian regime. 

The paper is structured as follows: We start by defining the model and summarize the available numerical results as well as the current status of the field-theoretical description of the quantum criticality in Sec.~\ref{sec:model} . In Sec.~\ref{sec:methods} we introduce the methods used to generate and evaluate the Monte Carlo data. This includes a brief review of the basic idea of the SSE QMC method (see Sec.~\ref{sec:sse}), the implementation of the long-range coupling on the finite lattice via Ewald sums (see Sec.~\ref{sec:ewald}), the protocol we used to extract zero-temperature results (see Sec.~\ref{sec:t0}), the observables we determined (see Sec.~\ref{sec:observables}) as well as finite-size scaling below and above the upper critical dimension (see Sec.~\ref{sec:fss}), followed by the details of the application of the scaling to our data in Sec.~\ref{sec:dataanalysis} and the extraction of the critical exponent $\nu$ in Sec.~\ref{sec:nu-yr}. Section~\ref{sec:results} contains the finite-size scaled results of the simulations of the ferromagnetic as well as the antiferromagnetic model on the one-dimensional chain as well as on the square lattice, including a discussion of our findings and comparisons with other available numerical data. Finally we conclude our work in Sec.~\ref{sec:conclusion}.

\section{Model} 

\label{sec:model}
We investigate the LRTFIM given by
\begin{align}
		\Hamil^{\text{LRTFIM}} = \frac{J}{2}\sum_{{\bf i} \neq {\bf j}} \frac{1}{|{\bf i}-{\bf j}|^{d+\sigma}}\sigma_{{\bf i}}^z\;\sigma_{{\bf j}}^z -h\sum_{{\bf j}}\sigma_{{\bf j}}^x  ~ , \label{eq:H_tfim_orig}
\end{align}
with Pauli matrices $\sigma_{{\bf i}}^{x/z}$ describing spins $1/2$ located on the lattice sites ${\bf i}$. The transverse field is tuned by the parameter $h>0$. The Ising interaction is ferromagnetic for $J<0$ and antiferromagnetic for $J>0$. The positive parameter $(d+\sigma)$ governs the decay of the coupling constants,
from a nearest-neighbor model for $\sigma=\infty$ to an all-to-all coupling for $d+\sigma=0$.
By separating the spatial dimension $d = 1, 2$ of the system from the decay exponent, we can treat one- and two-dimensional systems
on an equal footing. Let us note that some of the literature instead use the decay exponent $\alpha = d+\sigma$.

We investigate the Hamiltonian on finite linear chains of length $L$ and finite $L\times L$ segments of the square lattice. In both cases, we use periodic boundary conditions. The simulations are performed using the SSE QMC algorithm (see Sec.~\ref{sec:sse}). We focus on the limit $T\rightarrow0$ and $L\rightarrow \infty$, to contribute to the understanding of the quantum criticality of the LRTFIM.

There are several numerical studies concerning the quantum-critical properties of the LRTFIM on the one-dimensional spin chain \cite{Vodola2015,Fey2016,Zhu2018,Koffel2012,Sun2017,Vanderstraeten2018, Adelhardt2020}. For the antiferromagnetic chain, the established picture is that there is a phase transition of 2D Ising type between the high-field polarized phase and the low-field antiferromagnetic phase for all $\sigma\geq 1.25$ \cite{Fey2016,Sun2017,Koffel2012, Adelhardt2020}. 
In the regime of ultra-long-range couplings, recent finite-size density-matrix renormalization group findings suggest that the Ising universality holds for any $\sigma> -1$ \cite{Sun2017}. On the other hand, earlier results using matrix product states generalizing the time-dependent variational principle \cite{Koffel2012} indicate that critical exponents may vary for $\sigma<1.25$ and that the area law for the entanglement entropy may be violated. The stability of the low-field phase reduces for $\sigma\rightarrow -1$ \cite{Vodola2015,Fey2016,Koffel2012,Sun2017}.

The ferromagnetic chain shows a phase transition between the field-polarized and the ferromagnetic phase with three different regimes \cite{Fey2016,Zhu2018,Adelhardt2020}. First, for $\sigma>2$, the transition is of 2D Ising type with the well-known short-range exponents. For small $\sigma < 2/3$, the exponents are those of the long-range Gaussian theory \cite{Dutta2001}. 
In between, for $2/3 < \sigma < 2$, there is a regime of monotonously varying critical exponents that connects these two limits. Multiplicative logarithmic corrections to scaling are expected at the boundary of the long-range Gaussian regime ($\sigma = 2/3$) \cite{Dutta2001,Fey2016, Adelhardt2020}.

In two dimensions, the LRTFIM is numerically much more demanding.  The model has been studied on semi-finite triangular cylinder geometries \cite{Saadatmand2018,Koziol2019} and on the triangular lattice \cite{Humeniuk2016, Fey2019}. The LRTFIM on the square lattice as studied in this work has been investigated so far with high-order series expansions \cite{Fey2019}.  For the antiferromagnetic case, a 3D Ising phase transition between the polarized high-field phase and 2D antiferromagnetic order is found for all accessible $\sigma$. Very small $\sigma<0.5$ can not be well described by the high-order series expansion. The critical field $\hc$ is expected to vanish in the limit $\sigma \rightarrow -2$. In full analogy to the ferromagnetic chain, the square lattice with ferromagnetic interactions shows three regimes for the phase transition between the field-polarized phase and the ferromagnetic phase \cite{Fey2019}. For $\sigma > 2$, the square lattice shows 3D Ising nearest-neighbour criticality, while for $\sigma < 4/3$, the model is governed by long-range Gaussian theory
\footnote{Fey at al. mistakenly give the crossover to short-range behavior as $\sigma > 4$}.
In the intermediate regime, for $4/3 < \sigma < 2$, there are again continuously varying critical exponents and likewise for $\sigma = 4/3$ multiplicative logarithmic corrections to scaling are expected \cite{Dutta2001,Fey2019}.

The analogous three regimes for the ferromagnetic LRTFIM in one and two dimensions can be understood using a field-theoretical description of the quantum critical point \cite{Dutta2001,Defenu2017}. For ferromagnetic $d$-dimensional systems with algebraically decaying Ising interactions $1/|{\bf r}|^{d+\sigma}$ with distance $|{\bf r}|$ between the spins, the quantum critical point can be described by the real one-component quantum rotor action,
\begin{align}
		\label{eq:actionn}
		\mathcal{A}=\frac{1}{2}\int_{q,\omega} \, \tilde G_0^{-1}(q,\omega)\tilde \phi_{q,i\omega}\tilde \phi_{-q,-i\omega} + u\int_{x,\tau} \, \phi^4_{x,\tau}
\end{align}
using
\begin{align}
		\tilde G_0^{-1}(q,\omega)=\tilde g \omega^2+aq^\sigma+bq^2+r
\end{align}
with $a,b>0$ and $r, u$ being the mass and coupling term \cite{Dutta2001}. Note, the notation in Eq.~\eqref{eq:actionn} is taken from Ref.~\cite{Fisher1989}. For $\sigma \geq 2$ the leading terms in $q$ recover the nearest-neighbor $\phi^4$ Ising action of the $(d+1)$-dimensional Ising criticality \cite{Sachdev2007}. By scaling analysis of the Gaussian theory for $\sigma < 2$, it is possible to derive the long-range Gaussian critical exponents 
\begin{align}
		\gamma = 1, \hspace{0.75cm} \nu=\frac{1}{\sigma}, \hspace{0.75cm} \eta=2-\sigma, \hspace{0.75cm} z = \frac{\sigma}{2}.
\end{align}
The largest $\sigma$ for which the long-range Gaussian theory correctly describes the critical behavior is found from $d = \duppercrit(\sigma)$, where $d$ is the spatial dimensionality of the system and $\duppercrit$ is the upper critical dimension of the model (which depends on the decay exponent of the coupling).  Inserting the Gaussian exponents into the hyperscaling relation,
\begin{align}
		\label{eq:uppercritical}
		2-\alpha=\nu(d+z) \hspace{0.75cm} \xrightarrow[\nu=\frac{1}{\sigma}, z=\frac{\sigma}{2}]{\alpha=0}  \hspace{0.75cm} \duppercrit=\frac{3\sigma}{2} .
\end{align}
Dutta et al. \cite{Dutta2001} performed perturbative renormalization-group calculations around $\duppercrit$ and confirmed the stability of the Gaussian fixed point above the upper critical dimension. Therefore the Gaussian exponents hold for $\sigma < 2/3$ in one and $\sigma < 4/3$ in two dimensions. Using an $\epsilon$-expansion, perturbative corrections to the Gaussian exponents for $d<\duppercrit$ may be computed \cite{Dutta2001,Sachdev2007}. At the upper critical dimension multiplicative logarithmic corrections to scaling occur \cite{Fisher1972,Brezin1973,Wegner1973,Dutta2001}.

The precise behavior at the crossover to the short-range universality has been under debate for a long time \cite{Nagle1970,Fisher1972,Sak1973,Sak1977,Dutta2001,Behan2017,Defenu2017}.
Recently, it has been argued that the regime change is not exactly at $\sigma=2$, but at $\sigma=2-\eta_{\mathrm{SR}}$, with $\eta_{\mathrm{SR}}$ the anomalous dimension of the short-range transition \cite{Defenu2017}.

\section{Methods}\label{sec:methods}

\subsection{SSE Quantum Monte Carlo}
\label{sec:sse}
The SSE method we use is a finite-temperature QMC approach pioneered by A.~Sandvik to sample transverse-field Ising models with arbitrary (in particular long-range) interactions on arbitrary graphs \cite{Sandvik2003,Sandvik1991,Sandvik1992,Sandvik2010}. In this section we will briefly recapitulate and summarize the most important ideas in order to capture the essential aspects of the method.  For an in-depth understanding, we recommend Refs.~\cite{Sandvik2003,Sandvik2010}.

The starting point for the SSE approach is a representation of the Hamiltonian 
\begin{align}
	\Hamil=-\sum_{a,b}\Hamil_{a,b} 
\end{align}
as a sum of operators $\Hamil_{a,b}$. The choice of an orthonormal computational basis $\{\ket{\alpha}\}$ such that $\bra{\gamma}\Hamil_{a,b}\ket{\beta}\geq 0$ for all $\ket{\beta},\ket{\gamma}\in\{\ket{\alpha}\}$ avoids the sign problem.
Furthermore, the so-called no-branching rule
\begin{align}
		\Hamil_{a,b}\ket{\beta}\propto \ket{\gamma} \hspace{0.5cm} \forall{\ket{\beta}}\in \{\ket{\alpha}\} \hspace{0.3cm} \text{with} \hspace{0.3cm} \ket{\gamma} \in \{\ket{\alpha}\}
\end{align}
must be satisfied for the chosen decomposition of the Hamiltonian and the computational basis.
The partition function is rewritten as
\begin{align}
		\label{eq:sseseries}
		\myZ=\myTr\{e^{-\beta \Hamil}\}=\sum_{\alpha}\sum_{n=0}^\infty\sum_{\mySn}\frac{\beta^n}{n!}\bra{\alpha} \prod_{l=1}^n \Hamil_{a(l),b(l)} \ket{\alpha}
\end{align}
with $\mySn$ an ordered sequence of $n$ operator-index pairs
\begin{align}
		\mySn = \{[a(1), b(1)],[a(2),b(2)], ... , [a(n),b(n)])\} \ .
\end{align}
Two indices $a,b$ are commonly used for the LRTFIM \cite{Sandvik2003}\cite{Humeniuk2016}.

The SSE method approximates the series \eqref{eq:sseseries} by neglecting operator sequences $\mySn$ longer than some appropriately chosen $\mathcal{L}$. This leads to an efficient sampling scheme with an exponentially small error \cite{Sandvik2003}. Sequences with less than $\mathcal{L}$ operator-index pairs are padded to length $\mathcal{L}$ by randomly inserting identity operators. 
To compensate, one divides the partition funtion by the binomial coefficient $\mathcal{L}!/(n!(n-\mathcal{L})!)$ to obtain
\begin{align}
	\myZ & = \sum_\alpha\sum_{\mySL}\beta^n\frac{(\mathcal{L}-n)!}{\mathcal{L}!}\bra{\alpha} \prod_{l=1}^\mathcal{L} H_{a(l), b(l)} \ket{\alpha}
\notag\\
			 & = \sum_{\alpha}\sum_{\mySL} w ( \alpha, \mySL ).
\end{align}
The direct product $\{\ket{\alpha}\}\times\{\mySL\}$ is the configuration space of the SSE method.
Expectation values of observables are computed by Markov-chain Monte Carlo, in which the weights $w(\alpha, \mySL)$ describe the stationary distribution on the configuration space.  The weights are non-negative by virtue of the non-negative matrix elements of the operators $\Hamil_{a,b}$ constituting the Hamiltonian.

For the LRTFIM, it is suitable to use the $\sigma^z$-basis as the computational basis and to decompose the Hamiltonian as
\begin{align}
		\Hamil^{\text{LRTFIM}} &= -\sum_{a=0}^N\sum_{b=1}^N \Hamil_{a,b} - C \\
\intertext{where $C$ is a constant and}
		\Hamil_{0,0} &= \mathds{1}
\label{defIdentityOperator} \\
		\Hamil_{a,0} &= h\sigma_{\Gamma(a)}^x
\label{defFieldOperator} \\
		\Hamil_{a,a} &= h\mathds{1}
\label{defConstantOperator} \\
		\Hamil_{a,b} &= |J_{\Gamma(a),\Gamma(b)}|-J_{\Gamma(a),\Gamma(b)}\sigma_{\Gamma(a)}^z\sigma_{\Gamma(b)}^z.
\label{defBondOperator}
\end{align}
Here, $a, b>0$, and $a\not=b$ in Eq.~\eqref{defBondOperator}.
For the operator indices like $a$ or $b$ there is a bijective mapping $\Gamma$ to the corresponding lattice site ${\bf i}$. 

This configuration space is sampled according to the weights $w(\alpha, \mySL)$. The SSE method introduces two types of Monte Carlo moves. First, the diagonal updates replace the constant-field operators, \Eq{defConstantOperator}, and bond operators, \Eq{defBondOperator}, at sequence position $p$ with identity operators at the same position and vice versa,
\begin{align}
        [a,b]_p
        \leftrightarrow
        [0,0]_p
        \hspace{0.3cm}\text{with}\hspace{0.3cm} a,b \neq 0.
\end{align}
Second, the off-diagonal updates replace pairs of $[a,a]$ operators at positions $p_1$ and $p_2$ at lattice site $\Gamma(a)$ in the sequence with two transverse-field operators $[a,0]$,
\begin{align}
		[a,a]_{p_1}[a,a]_{p_2} \leftrightarrow [a,0]_{p_1}[a,0]_{p_2} \hspace{0.3cm}\text{with}\hspace{0.3cm} a \neq 0
\end{align}
and vice versa. Off-diagonal updates can be performed using local update schemes or quantum cluster updates \cite{Sandvik2003}.

In our simulations, one Monte Carlo sweep consists of two steps. First, we iterate through the operator sequence, performing the diagonal updates for each position $1,\dotsc, \mathcal{L}$ in the operator sequence. Each update is individually accepted or rejected with an acceptance probability satisfying detailed balance that results from the ratio of weights before and after the update \cite{Sandvik2003}. Second, for the off-diagonal update, we use Sandvik's quantum cluster update \cite{Sandvik2003}. The quantum cluster update identifies clusters in the Monte Carlo state in real space and imaginary time. Clusters are bounded in the imaginary-time direction by transverse-field operators \eqref{defFieldOperator} and constant-field operators \eqref{defConstantOperator}, but linked in real space by bond operators \eqref{defBondOperator}.  Clusters can be flipped by flipping the spin states and interchanging transverse-field and constant-field operators at the edge of the cluster at no change of weight. 
The two updates together suffice for the Monte Carlo algorithm to explore the entire $\{\ket{\alpha}\}\times\{\mySL\}$ configuration space \cite{Sandvik2003}.

\subsection{Long-Range Coupling}
\label{sec:ewald}
To optimally represent the thermodynamic limit on the finite-system simulation,
we use Ewald-corrected couplings \cite{Fukui2009,Humeniuk2018,Humeniuk2020}. In case of the one-dimensional chain, the corrected couplings have a closed form
\begin{align}
		\frac{1}{|i-j|^{1+\sigma}}& \longrightarrow \sum_{k=-\infty}^{\infty}\frac{1}{|i-j+kL|^{1+\sigma}} = \\
		&\frac{1}{L^{1+\sigma}} \left[ \zeta(1+\sigma, \frac{|i-j|}{L}) + \zeta(1+\sigma, 1 - \frac{|i-j|}{L})\right]\nonumber
\end{align}
using the Hurwitz $\zeta$ function
\begin{align}
		\zeta(s,q)=\sum_{n=0}^\infty\frac{1}{(q+n)^s} \hspace{0.5cm}\text{with}\hspace{0.5cm}\Re(s)>1 \wedge \Re(q)>0.
\end{align}
For the square lattice, we truncate the infinite series, and the couplings are
\begin{align}
		\frac{1}{|\mathbf i-\mathbf j|^{2+\sigma}}&\longrightarrow \sum_{k,l=-N_c}^{N_c}\frac{1}{|\mathbf i-\mathbf j+ kL \mathbf e_x + lL \mathbf e_y|^{2+\sigma}}
\end{align}
with $N_c$ being a cutoff value. We use $N_c=10^4$ for $\alpha\geq4$ and $N_c=10^5$ for $\alpha<4$.
By imposing the modified couplings we reduce finite-size effects due to neglecting interacting partners not included in the finite segment of the lattice.

\subsection{Simulation parameters for the zero-temperature sampling}
\label{sec:t0}
To find the temperature $T_{\text{eff},0}(L,h/J,\sigma)$ at which we effectively sample the zero-temperature properties of the system, we use an empirical scheme inspired by the beta-doubling method \cite{Sandvik2002}.
For two different linear system sizes $L_1>L_2$, one finds $T_{\text{eff},0}(L_1,h/J,\sigma)<T_{\text{eff},0}(L_2,h/J,\sigma)$ as the finite-size gap decreases with increasing system size. Therefore, a larger system size requires a lower simulation temperature. The finite-size finite-temperature crossover point is expected to scale as $L^z$ where $z$ is the dynamical correlation length exponent \cite{Sachdev2007,Kirkpatrick2015,Humeniuk2018}.
To ensure that we are sampling zero-temperature order parameter properties, we require the squared (staggered) magnetization curve of the largest system to converge in temperature. To probe this convergence, we successively halve the simulation temperature $T_{n}=\text{const.}\times\frac{1}{2^n}$.

To enhance convergence towards the ground-state space, we perform in the final simulations a stepwise cooling over all $T_n>T_{\text{eff},0}$.

\begin{figure}[ht]
	\centering
	\includegraphics{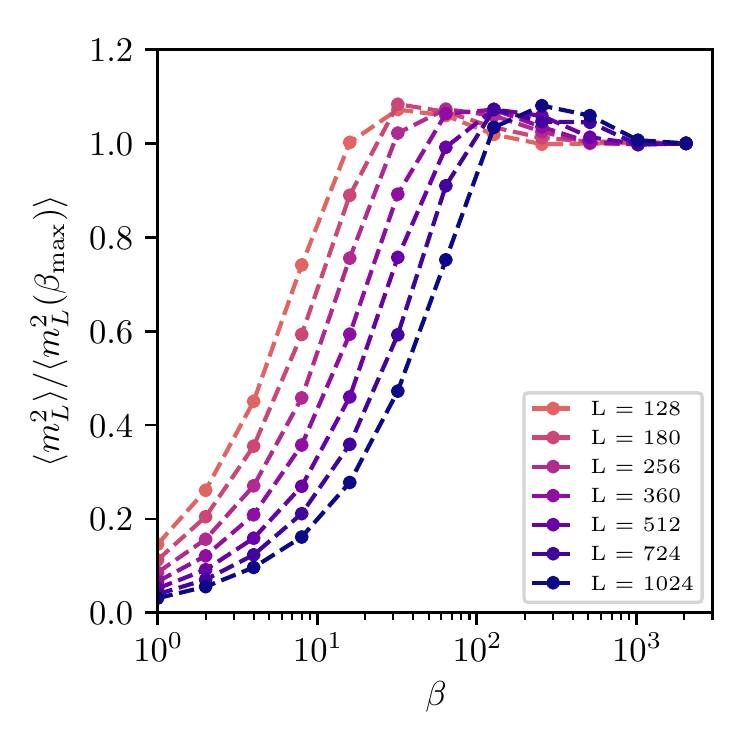}
	\caption{Behavior of the squared magnetization normalized by $\langle m^2_L(\beta_{\text{max}})\rangle$ with $\beta_{\text{max}}=2048$ for the \linchain\ in the short-range regime ($\sigma = 2.5$) with $z=1$, as the reciprocal temperature $\beta$ is increased. All systems larger than $L>1024$ were discarded from the analysis since they have not yet converged to the ground state of the system.
	}
	\label{fig:beta_doubling}
\end{figure}

\Fig{fig:beta_doubling} shows an example of this procedure.
We plot the squared magnetization for the ferromagnetic LRTFIM with $\sigma=2.5$ on the \linchain\ for different system sizes $L$. We choose a transverse-field value of $h=1.25$ in the vicinity of the quantum critical point ($\hc=1.25001(2)$) as we are interested in the convergence of the magnetization at criticality, and because the finite-size gap should be the smallest at criticality. 
As expected from theory, the magnetization converges faster to the zero-temperature limit $\beta\rightarrow\infty$ for smaller system sizes,
in accordance with the scaling $L^{z=1}$.

\subsection{Observables}
\label{sec:observables}

As the static properties of a second-order phase transition are investigated, the main focus lies on the examination of the order parameter, the (staggered) magnetization of the symmetry-broken phase,
\begin{align}
    m_{L,{\bf q}} = \frac{1}{L^d}\sum_{{\bf j}} \sigma^z_{\mathbf j} \eexp^{\mathrm i{\bf q}\cdot {\bf j}}.
\end{align}
The ordering momentum $\mathbf q$ of the symmetry-broken phase is ${\bf q}=(0,0)$ for ferromagnetic couplings and \mbox{${\bf q}=(\pi,\pi)$} for antiferromagnetic couplings on the square lattice (${\bf q}=0$ and ${\bf q}=\pi$ on the \linchain). We define the mean magnetization for a system of linear system size $L$ as $M_{L,{\bf q}}=\langle m_{L,{\bf q}}\rangle $.
We will extract critical exponents from the data collapse of the squared magnetization $\expect{m^2_{L,{\bf q}}}$.

To locate the phase transition we also calculate the Binder cumulant \cite{Binder1981,Binder1981a,Binder1987}
\begin{align}
    U_{L, {\bf q}} =\frac{3}{2}\left(1-\frac{\langle m^4_{L,{\bf q}}\rangle}{3\langle m^2_{L,{\bf q}}\rangle^2}\right).
\label{defBinderCum}
\end{align}
Due to the scaling behavior of the magnetization at the critical point, the Binder cumulant becomes independent of the system size up to corrections (see Sec.~\ref{sec:fss}).
Therefore, the intersection points of the Binder cumulants for different $L$ can be used to locate the critical point.
The prefactors in \Eq{defBinderCum} are chosen such that $U_{L,{\bf q}}\rightarrow 1$ in the ordered phase and $U_{L,{\bf q}}\rightarrow 0$ in the paramagnet.

In the following, we will always set $\vec q$ to the ordering momentum of the system, and drop it from the notation.

\subsection{Finite-size scaling below and above the upper critical dimension}
\label{sec:fss}

We rely on the scaling hypothesis \cite{Widom1965,Domb1965,Kadanoff1966,Patashinskii1966,Wilson1971,Stanley1999} to extract infinite-system critical properties from our finite simulations. We assume that an observable $\Observ$ with a divergence $\Observ (r, T=0)\sim |r|^{\omega}$ satisfies the scaling relation
\begin{align}
		\label{eq:homscaling}
		\Observ(r,T=0)=b^{-\omega y_r}\Observ(rb^{y_r},T=0)
\end{align}
in the vicinity of the phase transition \cite{Kirkpatrick2015,Sachdev2007},
with $r=|h/\hc-1|$ the distance from the critical point, $\omega$ the critical exponent of $\Observ$, $b$ the scaling parameter and $y_r=1/\nu$ the scaling power of $r$.
We are however investigating finite systems such that the observable $\Observ_L(r, T=0)$ we measure depends on the linear system size $L$. Following the argumentation of Ref.~\cite{Cardy2012} including $L^{-1}$ as a variable of the scaling form with a scaling power of one and using $b=L/L_0$ one finds
\begin{align}
		\Observ_L(r, T=0) &= (L/L_0)^{-\omega y_r}\Observ_{L_0}(r(L/L_0)^{y_r}, T=0)\nonumber\\
		\label{eq:scaling-form}
		& = L^{-\omega y_r}\Psi_{\Observ}(rL^{y_r}, T=0),
\end{align}
where we introduced the $L$-independent scaling function $\Psi_{\Observ}$ and absorbed $L_0$ into its definition.
\Eq{eq:scaling-form} is the finite-size scaling (FSS, \cite{Fisher1972a,Brezin1981,Brezin1985,Binder1981,Binder1987}) form.

From the field theory of the ferromagnetic LRTFIM it is known that the upper critical dimension is lowered by decreasing $\sigma$ \cite{Dutta2001}.
Thus, by Eq.~\eqref{eq:uppercritical}, there is a $\sigma^*$ at which $d_{\mathrm{uc}}$ becomes equal to the physical dimension $d$.
For $\sigma < \sigma^*$, hyperscaling is no longer valid and the phase transition is described by the long-range Gaussian theory \cite{Dutta2001}. In this regime the $\phi^4$-coupling $u$ becomes a dangerous irrelevant variable \cite{Privman1988,Binder1985,Binder1987,Berche2012,Kenna2014a,Kenna2014b,Flores2015}.

When dangerous irrelevant variables like the coupling $u$ with a scaling power $y_u$ are present these have to be taken into account in generalized homogeneous scaling functions and are usually treated by modifying the other scaling powers \cite{Binder1985,Binder1987,Kirkpatrick2015} such that \Eq{eq:scaling-form} becomes
\begin{align}
		\label{eq:scaling-form*}
		\Observ_L(r, T=0)=L^{-\omega y_r^*}\Psi_{\Observ}(rL^{y_r^*}, T=0).
\end{align}
The modified scaling power $y_r^*=y_r+p_r y_u$ absorbs the effect of dangerous irrelevant variables and it does not coincide with the reciprocal correlation length exponent anymore.
Historically, the main approach to such problems was to introduce another length scale $l$ with a divergence $l\sim |r|^{-1/y_r^*}$ replacing the correlation length $\xi \sim |r|^{-1/y_r}$ as the characteristic length scale of the system \cite{Binder1985,Binder1987}.

A more recent approach developed for thermal phase tansitions is called Q finite-size scaling (Q-FSS) \cite{Berche2012,Kenna2014a,Kenna2014b,Flores2015}.
The key idea is to include the effect of dangerous irrelevant variables to the correlation sector and from this allow the finite-size correlation length to scale as \mbox{$\xi_L \sim L^\koppa$}, where $\koppa=y_r^*/y_r$ above $\duppercrit$ and $\koppa=1$ otherwise \cite{Berche2012,Kenna2014a,Kenna2014b,Flores2015}.
By imposing the generalized scaling of the correlation length the scaling form is modified to
\begin{align}
		\label{eq:qfss}
		\Observ_L(r)=L^{-\omega \koppa/\nu}\Psi_{\Observ}(r L^{\koppa/\nu}) \ .
\end{align}
Note, here $r$ is the tuning parameter of the thermal phase transition.
For classical systems Q-FSS predicts $\koppa = \max(1,d/\duppercrit)$ \cite{Berche2012,Kenna2014a}. Unfortunately, the same line of reasoning does not directly apply to the quantum case.
In order to avoid confusion with the classical case, we use for the LRTFIM
\begin{align}
		\label{eq:scaling-form-t}
		\Observ_L(r, T=0)=L^{-\omega \tilde{y}_r}\Psi_{\Observ}(rL^{\tilde{y}_r}, T=0),
\end{align}
by introducing a generalized scaling power
\begin{align}
	\tilde{y}_r = 
	\begin{cases}
		y_r \text{ for d < \duppercrit}\\
		y_r^* \text{ for d > \duppercrit}
	\end{cases}
\end{align}
and therefore unifying Eqs.~\eqref{eq:scaling-form} and~\eqref{eq:scaling-form*}.

\subsection{Extraction of critical properties}
\label{sec:dataanalysis}

The critical transverse field \hc\ is most conveniently extracted from the crossing points of the Binder cumulants,
\Eq{defBinderCum}, for different system sizes \cite{Binder1987}.
The Binder cumulant $U$ is an observable with a scaling power of zero, \ie,
\begin{align}
	U_L (r, T = 0) = \Psi_U(rL^{\tilde{y}_r}, T = 0),
\end{align}
and thus becomes independent of $L$ for $r=0$ to leading order.
Thus, pairwise intersections of $U_L(h)$ and $U_{bL}(h)$ for different system sizes $L$ and $bL$ ($b>1$) give an estimate for $\hc$. These intersections can be extrapolated for families of $L$ towards $b\rightarrow \infty$ in order to take into account leading corrections to scaling \cite{Binder1987}. 
Similar extrapolation methods for critical exponents are described in \ Ref.~\cite{Binder1987}. For critical properties we find a self-consistent data collapse of squared (staggered) magnetization curves to be more stable than these extrapolation methods.
In our simulations, the increased stability outweighs the value of including corrections.
All the results reported in the paper come from the data collapse method described in the following.

In the vicinity of the critical point, due to the generalized FSS form \Eq{eq:scaling-form-t}, the transformations
$
	r \rightarrow rL^{\tilde{y}_r}
$ and $
	\Observ_L \rightarrow \Observ_L L^{\omega \tilde{y}_r}
$
will collapse the data $\Observ_L(r,T~=~0)$ for different system sizes onto the scaling function.
We use this behavior to determine $\tilde{y}_r$, $\omega$, and $\hc$.
In practice, we achieve this by fitting the whole set of data points $\Observ(X)$ with $X=(L,h)$
for all $L$ simultaneously to the function
\begin{align}
		\label{eq:2dfss}
		F(X) &= L^{-\omega \tilde{y}_r}p\left[ (h-\hc)L^{\tilde{y}_r}\right]
\end{align}
where $p(x) = a_0 + a_1 x + \dotsm + a_n x^n$ is a polynomial with coefficients $\vec a = (a_0, a_1, \dots, a_n)$
approximating the scaling function.
For the data collapse we used the mean squared magnetization $\expect{m^2_{L,{\bf q}}}$ such that $\omega = 2\beta$ and the fit thus determines the parameter set $\hc, \tilde{y}_r, \beta$ and $ \vec{a}$. Note that we can determine $\beta$ independently of $\tilde{y}_r$ and therefore $\beta$ can be directly extracted both below and above the upper critical dimension. We discuss the extraction of $\nu$ based on $\tilde{y}_r$ in Sec.~\ref{sec:nu-yr}.

In principle, \hc\ is already determined to good accuracy from the crossing points of the Binder cumulants.
However, in the known limits of short-range and Gaussian critical exponents, we find better values for the exponents when \hc\ is a free fit parameter than when we fix the value from the Binder cumulant crossings.
This shows that the fit is very sensitive to the precise value of \hc.
On this basis, we present the results from fitting with free \hc.

\subsection{Extraction of $\nu$ from $\tilde{y}_r$}
\label{sec:nu-yr}
From the data collapse of the calculated observables we are able to obtain an estimate for the critical point $\hc$, the critical exponent $\omega$ of the respective observable as well as the scaling power $\tilde{y}_r$ of $r$ in the generalized homogeneous scaling function.
Below the upper critical dimension, $\tilde{y}_r= y_r = 1/\nu$ and we obtain $\nu$ directly from the data collapse. Above the upper critical dimension $\tilde{y}_r$ no longer coincides with the reciprocal correlation length exponent and it is a priori not clear how to extract $\nu$.

\begin{figure}[ht]
	\centering
	\includegraphics{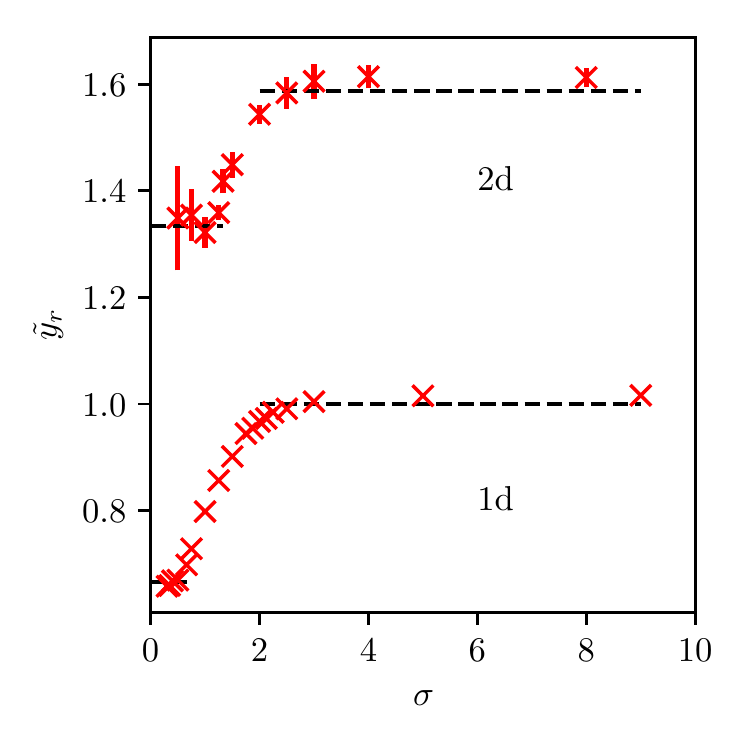}
	\caption{Scaling power $\tilde{y}_r$ for the ferromagnetic LRTFIM on the \linchain\ (1d) and \sqlattice\ (2d) extracted from the self-consistent data collapse of the mean squared magnetization. In both dimensions $\tilde{y}_r$ appears to level off above the upper critical dimension.}
    	\label{fig:yt}
\end{figure}

In \Fig{fig:yt} we present the extracted values for $\tilde{y}_r$. In both dimensions the scaling power seems to level off at the value of the scaling power at the upper critical dimension $\tilde{y}_r = \sigma_{\text{uc}}$ predicted from the long-range Gaussian theory.
We therefore find for the ratio $y_r^*/y_r = \sigma_{\text{uc}}/\sigma = d/\duppercrit $ above the upper critical dimension. 
It is noteworthy that even though there are subtle differences to the classical Q-FSS approach \cite{Berche2012,Kenna2014a,Kenna2014b,Flores2015} due to the appearence of $z$ in the scaling form of the free energy, we arrive at the same ratio $y_r^*/y_r= d/\duppercrit$ for the scaling powers.

With this we are able to obtain estimates for $\nu$ in the long-range Gaussian regime through 
\begin{align}
	\label{eq:conversion}
	\nu = \frac{d}{\duppercrit}\frac{1}{y_r^*}.
\end{align}
We are aware that a good agreement with the predictions from the long-range Gaussian theory is to be expected as the conversion is constructed such that it fulfills these expectations. Nevertheless, the consistency between both dimensions as well as the analogy to the Q-FSS support this proceeding.

\section{Results and discussion}
\label{sec:results}

\subsection{Ferromagnetic coupling}
\begin{figure*}[p]
	\centering
	\includegraphics{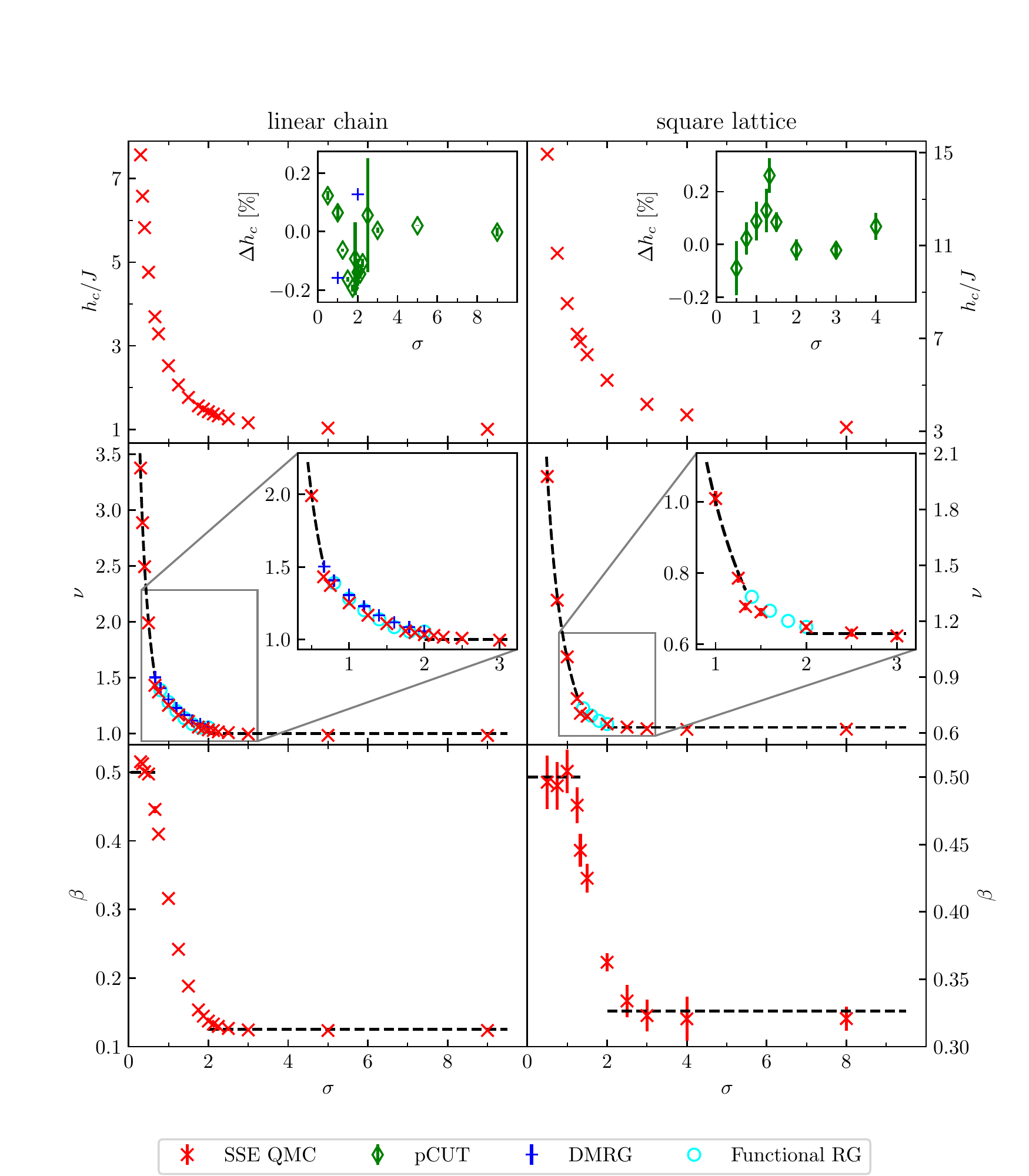}
    \caption{Critical field values $h_c$ and critical exponents $\nu$, $\beta$ for the ferromagnetic phase transition on the \linchain\ (left column) and the \sqlattice\ (right column).
        Top row: The critical field \hc.
		The inset shows the relative difference to the high-field series expansion pCUT results from \Refs{Fey2016,Fey2019} (green diamonds) and to the density-matrix renormalization group results from Ref.~\cite{Zhu2018} (blue pluses). We define $\Delta\hc = 100\%\cdot (\hc-\hc^{\mathrm{ref}})/\hc$.
        Center and bottom row: The critical exponents $\nu$ and $\beta$, respectively.
		The dashed lines depict the expected nearest-neighbor values in the large $\sigma$ regime \cite{Pfeuty1971,Kos2016} and long-range Gaussian exponents at small $\sigma$ values. 
		For the $\nu$ values in one dimension we also present the density-matrix renormalization group results from Ref.~\cite{Zhu2018} (blue pluses) and in one and two dimensions the functional renormalization group results from Ref.~\cite{Defenu2017} (cyan circles).
    }   
    \label{fig:ferroresults}
\end{figure*}

The ferromagnetic LRTFIM behaves qualitatively the same on the \linchain\ and on the \sqlattice\ (see Sec.~\ref{sec:model}). We present the critical-field values $\hc$ and critical exponents $\beta$ and $\nu$ for both dimensionalities in Fig.~\ref{fig:ferroresults}.

The top panels of \Fig{fig:ferroresults} show that the critical field $\hc$ diverges as $\sigma\rightarrow 0$.
This behavior is intuitively explained by the fact that a decreasing $\sigma$ increases the excitation cost in the low-field phase. On the other hand, a smaller $\sigma$ increases the mobility of the high-field excitations, therefore leading to a higher $\hc$.
Over the whole range of $\sigma$, our results are in very good agreement with the findings from high-field series expansions \cite{Fey2016,Fey2019} (see also insets in Fig.~\ref{fig:ferroresults}), with differences of around $0.1\%$. We find that the SSE method reaches the same level of accuracy as the high-field expansion.
The one-dimensional $\hc$ values are also consistent with the findings of \Ref{Zhu2018}, also shown in the inset.

The center and bottom panels of \Fig{fig:ferroresults} show the critical exponents $\nu$ and $\beta$.
Both in the nearest-neighbor regime and in the long-range Gaussian regime, the exponents agree very well with field theory predictions. In the short-range regime, exponents are systematically underestimated by $1-2\%$ as compared to the known 2D/3D-Ising short-range exponents. These deviations are due to slow logarithmic convergence of the finite-size scaling. Note that we observe a similar shift for the antiferromagnetic systems.

In the intermediate regime, the critical exponents increase monotonously as $\sigma$ decreases.
For even lower values of $\sigma$, the exponents track the predictions from long-range Gaussian field theory with high fidelity.
In the long-range regime, the dynamical critical exponent vanishes as \mbox{$z = \sigma/2$}, and thus the finite-size gap decreases more slowly. Somewhat counterintuitive, this allows us to access larger system sizes in the long-range regime than in the $z=1$ nearest-neighbor case.

The boundaries between the three different regimes are not sharp.
Reasons for this include insufficient system sizes as well as multiplicative logarithmic corrections \cite{Fisher1972,Dutta2001} at the upper critical dimension that we have not attempted to correct for.
In one dimension, we were able to simulate larger linear system sizes, and find considerably less rounding and shifting of the boundaries between the regimes.
A comparison with data from available functional renormalization group \cite{Defenu2017} shows a similar behavior in the intermediate regime, but we were not able to resolve the shift from $\sigma =2$ to $\sigma = 2-\eta_\mathrm{SR}$ due to the rounding of the boundary.

Finally, we stress that the conversion between the scaling power $\tilde y_r$ and the correlation length exponent $\nu$ as shown in \Eq{eq:conversion} appears to be applicable to the quantum phase transition of the LRTFIM and finds the correct Gaussian exponents above the upper critical dimension.

\subsection{Antiferromagnetic coupling}

\begin{figure*}[p]
    \centering
    \includegraphics{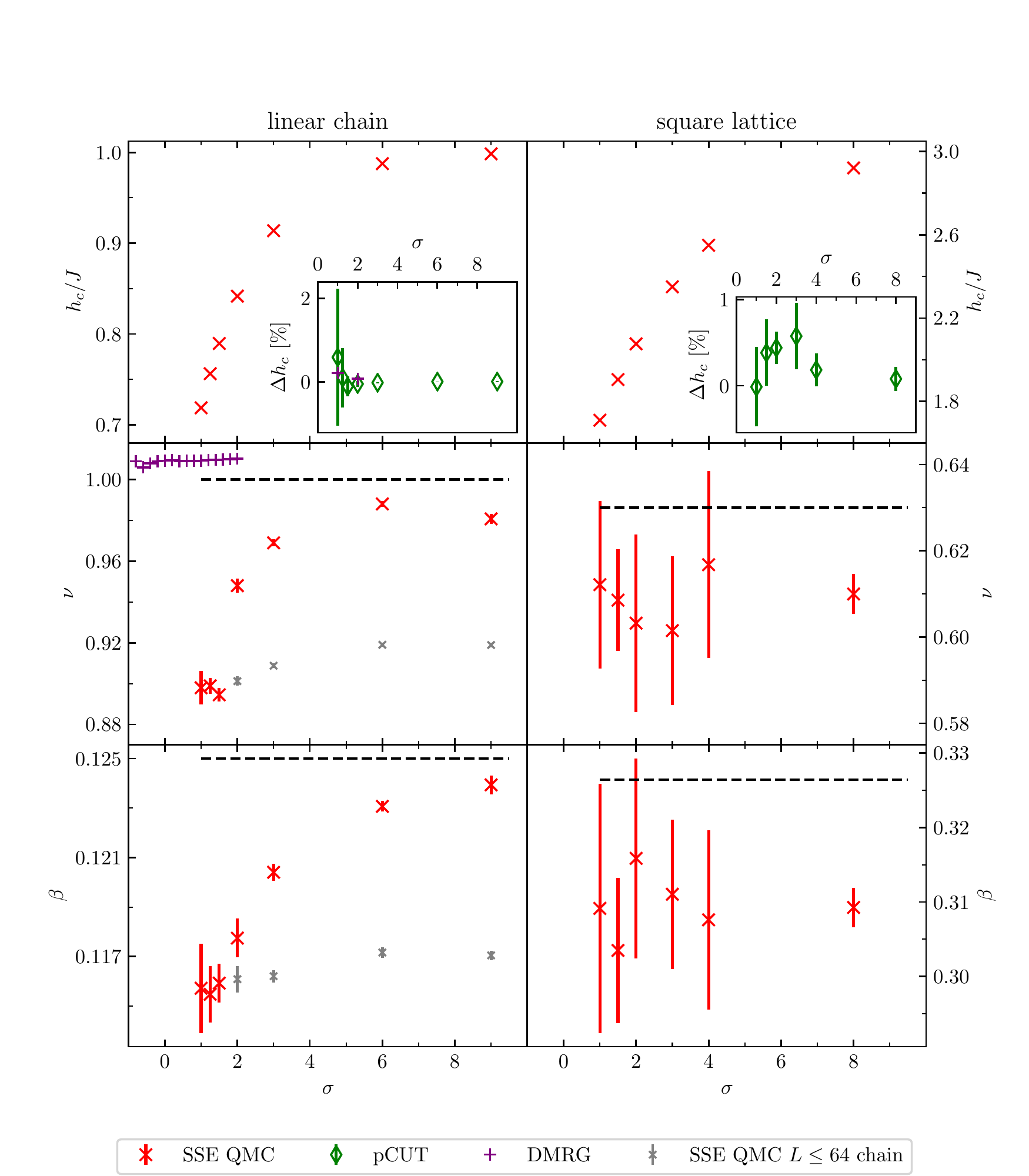}
    \caption{Critical field values $h_c$ and critical exponents $\nu$, $\beta$ for the antiferromagnetic phase transition on the \linchain\ (left column) and the \sqlattice\ (right column).
        Top row: The critical field \hc.
			The inset shows the relative difference to the high-field series expansion pCUT results from \Refs{Fey2016,Fey2019} (green diamonds) and to the density-matrix renormalization group results from Ref.~\cite{Sun2017} (purple pluses). We define $\Delta\hc = 100\%\cdot (\hc-\hc^{\mathrm{ref}})/\hc$.
        Center and bottom row: The critical exponents $\nu$ and $\beta$, respectively.
            Dashed lines indicate the known critical exponents for the short-range 2D and 3D Ising model \cite{Pfeuty1971,Kos2016}.
			For the one-dimensional chain we also present the $\nu$ values from Ref.~\cite{Sun2017} (purple pluses) as well as values for $\nu$ and $\beta$ obtained by collapsing curves of the same small system sizes as for the three smallest $\sigma$ (small gray crosses).
    }   
    \label{fig:afresults}
\end{figure*}

The top row of \Fig{fig:afresults} depicts the critical field values for the phase transition in the antiferromagnetic LRTFIM,
between the x-polarized high-field phase and the symmetry-broken antiferromagnetic low-field phase.
The critical field \hc\ decreases for decreasing $\sigma$, in agreement with previous high-field series expansion pCUT results \cite{Fey2016, Fey2019}.
Quantitatively, pCUT predicts somewhat smaller critical fields than our SSE results.
The pCUT expansion converges better at large fields; at low fields, the DlogPadé extrapolation tends to underestimate the critical field \hc.
We thus expect that the antiferromagnetic \hc\ obtained from SSE are more accurate than the pCUT results.
We are not aware of other previous results for the 2D system.
For the linear chain, the other available $h_c$ values \cite{Sun2017, Koffel2012, Vodola2015} are in good agreement with our results where applicable.
For very small $\sigma < 1$, the auto-correlation times of the SSE Monte Carlo dynamics increase as more and more bond operators are present in the operator sequence and the field operators diffuse only slowly.
In this regime, we cannot extract critical exponents from SSE simulations.

The center and bottom rows of \Fig{fig:afresults} display our results for the critical exponents $\nu$ and $\beta$, respectively.
Current literature does not predict a change from the universality class of the nearest-neighbor transverse-field Ising model for $\sigma\geq 1.25$. Possible deviations from the nearest-neighbor criticality are only reported for the one dimensional chain \cite{Koffel2012,Vodola2015}. Following the findings of \Refs{Fey2016,Sun2017,Fey2019}\ we expect to find 2D and 3D Ising critical exponents (dashed lines, \Refs{Pfeuty1971,Kos2016}) over the investigated range of $\sigma$.

On the square lattice (right column in \Fig{fig:afresults}), for both $\nu$ and $\beta$, our SSE results are systematically below the short-range exponents by about $3-6\%$.
As in the ferromagnetic system, this systematic shift towards smaller critical exponents is present even at large $\sigma$, where the nearest-neighbor criticality must be recovered (see above). Again, we conclude that this shift represents slow convergence of the finite-size scaling.
For all antiferromagnetic two-dimensional systems, we used the same maximum system size of $N=900$ spins.

On the \linchain, we observe a somewhat smaller shift of $2-3\%$ from the expected short-range values for $\sigma > 4$ (left column in \Fig{fig:afresults}).  Systems of $L=N=1024$ spins are accessible in this range.
For longer-range interactions, we experience greater difficulty:  The antiferromagnetic chain has a smaller finite-size gap and requires much lower temperatures in the SSE simulation to reproduce the ground-state physics.
This leads to a strong increase of the auto-correlation times of the Monte Carlo dynamics and to poor data quality.
For the three smallest $\sigma$ values considered, systems no larger than $L=64$ can be accessed.
We find that the underestimation of critical exponents worsens in this range, consistent with the hypothesis that the systematic error with respect to short-range exponents is due to insufficient system sizes.
Indeed, when we constrain the maximum system size to $L\leq 64$ for all $\sigma$, the critical exponents are underestimated
by the same magnitude over the whole range of $\sigma$ (see the small grey crosses for the \linchain\ in \Fig{fig:afresults}).
Thus, our findings are compatible with the short-range Ising universality classes for $\sigma>1$ both in one and two dimensions.

\section{Conclusion}
\label{sec:conclusion}
We study the LRTFIM on the \linchain\ and the \sqlattice\ using SSE QMC, both for ferromagnetic and antiferromagnetic Ising interactions,
to investigate static properties of the zero-temperature quantum criticality.
We extract the critical fields $h_c$ and the critical exponents $\beta$ and $\nu$ by means of finite-size scaling.
Above the upper critical dimension, we use a conversion inspired by the recent Q-FSS approach \cite{Berche2012,Kenna2014a,Kenna2014b,Flores2015}.
The obtained critical values \hc\ match state-of-the art series expansion values \cite{Fey2016,Fey2019} and are consistent with other available values in the literature \cite{Koffel2012,Vodola2015,Sun2017,Zhu2018}.
For the ferromagnetic LRTFIM, the calculated exponents $\beta$ and $\nu$ display the expected three regimes very well in both dimensions. There is a slight mismatch at the borders of the intermediate regime, but besides these minor deviations caused by 
multiplicative corrections to scaling at the upper-critical dimension and probably the systematic limitations to finite systems the main features of the critical properties are captured well by the finite-size scaling analysis.

In the antiferromagnetic case increasing auto-correlation times make it hard to simulate for small $\sigma$. This impacts the one-dimensional chain more than the square lattice, as lower temperatures are necessary to obtain effective zero-temperature results due to the larger linear system sizes simulations. For the two-dimensional square lattice we find within error bars the same critical exponents throughout the entire investigated $\sigma>1$ range with systematic errors of about $3\%-6\%$ deviation to the nearest-neighbor critical values. As we find the same exponents in the known limit $\sigma\rightarrow \infty$ makes us confident that the nearest-neighbor criticality spans across the investigated range. 

Regarding the antiferromagnetic LRTFIM, algorithmic developments in order to tackle the competing interactions will become necessary.
Overall, the interplay between geometrical frustration and long-range interactions as for the LRTFIM on the triangular lattice represents a major challenge and certainly deserves further attention in future research.

\section{Acknowledgments}
We thank Sebastian Fey for providing the pCUT data from references \cite{Fey2016,Fey2019}. We gratefully acknowledge the computational resources and support provided by the HPC group of the Erlangen Regional Computing Center (RRZE).
KPS acknowledges the support by the German Research Foundation (Project-ID 429529648 - TRR 306 QuCoLiMa „Quantum Cooperativity of Light and Matter“).

\bibliography{bibliography}

\end{document}